\documentclass[11pt,hidelinks]{article}
\pdfoutput=1
\usepackage[body={17.5cm, 22cm},right=2cm]{geometry}
\usepackage{color}
\usepackage{amssymb,graphicx}
\usepackage{amsmath}
\usepackage{epsfig}
\usepackage[bookmarks=true,pdfborder={0 0 0}]{hyperref} 
\usepackage{cite}
\usepackage{booktabs}
\allowdisplaybreaks

\begin{document}
\setcounter{page}{0}
\thispagestyle{empty}

\parskip 3pt

\font\mini=cmr10 at 2pt

\begin{titlepage}
~\vspace{1cm}
\begin{center}

{\LARGE \bf Topological Susceptibility and QCD Axion Mass: \\ [10pt]
QED and NNLO corrections }

\vspace{1cm}

{\large
Marco Gorghetto$^{a}$ and Giovanni~Villadoro$^{b}$}
\\
\vspace{.6cm}
{\normalsize { \sl $^{a}$ 
SISSA International School for Advanced Studies and INFN Trieste, \\
Via Bonomea 265, 34136, Trieste, Italy }}

\vspace{.3cm}
{\normalsize { \sl $^{b}$ Abdus Salam International Centre for Theoretical Physics, \\
Strada Costiera 11, 34151, Trieste, Italy}}

\end{center}
\vspace{.8cm}
\begin{abstract}
We improve the precision of the topological susceptibility of QCD, 
and therefore of the QCD axion mass, by including $O(\alpha_{\rm em})$ 
and NNLO corrections in the chiral expansion, which amount to 0.65(21)\% 
and -0.71(29)\% respectively. Both corrections are
one order of magnitude smaller than the known NLO ones, confirming 
the very good convergence of the chiral expansion and its reliability.
Using the latest estimates for the light quark masses the current 
uncertainty is dominated by the one of the low-energy constant 
$\ell_7$. When combined with possible improvements on 
the light quark mass ratio and $\ell_7$ from lattice QCD, 
our computation could allow to determine the
QCD axion mass with per-mille accuracy.
\end{abstract}

\end{titlepage}

\tableofcontents

\section{Topological Susceptibility at LO and NLO}\label{sec:LONLO}
In QCD the topological susceptibility $\chi_{\rm top}$ is one of the fundamental observables
describing the non-trivial properties of the QCD vacuum. Defined as the second derivative of
the free energy with respect to the $\theta$-angle at $\theta=0$, it determines how the QCD vacuum energy depends on the CP-violating $\theta$ parameter. While $\chi_{\rm top}$ 
vanishes at all order in perturbation theory, at high temperatures its value is expected 
to be well reproduced by semi-classical small-instanton configurations.\footnote{
Several lattice simulations in pure Yang-Mills~\cite{Berkowitz:2015aua,Borsanyi:2015cka} and in QCD \cite{Petreczky:2016vrs,Borsanyi:2016ksw,Burger:2018fvb,Bonati:2018blm} 
indicate that the temperature behavior predicted by
the instanton gas approximation might be valid already at small temperatures, just above the QCD
transition, although the overall size of $\chi_{\rm top}$ is not yet well reproduced.} 
At zero temperatures such a description is not valid, and indeed in the chiral limit current algebra relates $\chi_{\rm top}$ to the chiral condensate, which is notoriously associated to renormalons rather than to a semi-classical field configuration.

Recently the determination of $\chi_{\rm top}$ has seen a wave of renewed interest. Indeed,
the most plausible known solution to the strong-CP problem \cite{Peccei:1977hh} 
involves the presence of 
a light pseudo-Goldstone boson, the QCD axion~\cite{Weinberg:1977ma,Wilczek:1977pj,
Kim:1979if,Shifman:1979if,Zhitnitsky:1980tq,Dine:1981rt}, 
whose mass is determined by $\chi_{\rm top}$ through the relation\footnote{Corrections to 
this formula are of order $m_\pi^2/f_a^2$, which are negligible given that $f_a\gtrsim 10^8$~GeV 
\cite{Tanabashi:2018oca}.} 
$m_a^2=\chi_{\rm top}/f_a^2$ (where $f_a$ is the Peccei-Quinn scale 
controlling the axion coupling to the Standard Model). 
Since the existence of the QCD axion could also explain the dark matter abundance in our Universe~\cite{Preskill:1982cy,Abbott:1982af,Dine:1982ah}, a multitude of experiments are being pursued to search for this particle (see e.g.~\cite{Graham:2015ouw,Irastorza:2018dyq}). 
Using various forms of resonance effects to amplify the otherwise too feeble signal, several
of these experiments would be able to measure the axion mass with very high precision,
even down to ${\cal O}(10^{-6})$. When combined with measurements of the axion couplings and possibly the information of the axion relic abundance, such precision 
could be used to learn about the dynamics of new physics at much higher scales, 
as well as physics of the early universe including inflation, reheating and pre-BBN evolution.

At a first look a high precision determination of $\chi_{\rm top}$ 
seems hampered by its non-perturbative nature. On the contrary, chiral effective theories are particularly powerful in this case. By exploiting the freedom to rotate the whole $\theta$-dependence of the QCD functional into the phases of the lightest quark masses, the $\theta$-dependence of low-energy QCD observables (e.g. the vacuum energy) can be computed analytically using chiral Lagrangians, which are expansions in the 
light quark masses. In particular using the lightest two quarks, 
the effective expansion parameter, $m_{u,d}/m_s \sim$~few percent, is rather small, 
suggesting a fast convergence. 

Indeed, as shown explicitly in~\cite{diCortona:2015ldu}, the leading order formula~\cite{Weinberg:1977ma} 
\begin{equation} \label{eq:chiLO}
\chi_{\rm top}^{\rm LO}=\frac{z}{(1+z)^2}m_{\pi^0}^2 f_\pi^2 
\,, \qquad z\equiv \frac{m_u}{m_d}\,,
\end{equation}
where $m_{\pi^0}$ is the physical neutral pion mass and $f_\pi$ its decay constant (normalized as $f_\pi\sim 92.3$ MeV), is accurate at the few percent level. Next to leading order (NLO) corrections in the
chiral expansion have been computed in~\cite{diCortona:2015ldu}, and result in
\begin{align} \label{eq:chiNLO}
\chi^{\rm NLO}_{\rm top}=&\ \chi_{\rm top}^{\rm LO}\left[1+\delta_1\right] \,, \nonumber \\
\delta_1=&\ 2\frac{m_{\pi^0}^2}{f_\pi^2} \left[ h^r_1-h^r_3-\ell^r_4 - 
	(1-2 \Delta^2)\ell_7 \right] \,, \qquad \Delta\equiv\frac{1-z}{1+z}\,.
\end{align}
where the coefficients $h_i^r$ and $\ell_i^r$ are the low-energy constants (LECs) of the chiral Lagrangian defined in ref.~\cite{Gasser:1983yg}.
Using the available estimates for the quark mass ratio $z=0.48(3)$ and for the LECs, the topological susceptibility and the corresponding axion mass were
estimated to be
\begin{equation} \label{eq:chiold}
\chi_{\rm top}^{1/4}=75.5(5)~{\rm MeV}\,, \qquad m_a = 5.70(6)_z(4)_{\ell_i^r}\,\mu{\rm eV}\,\frac{10^{12}~{\rm GeV}}{f_a}\,,
\end{equation}
where the uncertainties in $m_a$ come respectively from $z$ and the LECs.
This estimate represents the current state of the art for the topological susceptibility and
the QCD axion mass.

Since the results in ref.~\cite{diCortona:2015ldu} new lattice simulations became available. 
Direct measurements of the topological susceptibility were performed 
in the isospin limit $z=1$ in refs.~\cite{Bonati:2015vqz} and \cite{Borsanyi:2016ksw}, which, 
once corrected for the leading isospin-breaking effects (i.e. the factor 
$z^{1/4}\sqrt{2/(1+z)}$), give respectively $\chi^{1/4}_{\rm top}= \,$75(3) and 75(2)~MeV, 
in nice agreement with eq.~(\ref{eq:chiold}).
While the current precision of these estimates is still roughly a factor of four worse than 
(\ref{eq:chiold}), the situation may change in the near future as systematics are further reduced on the lattice side. 

At the same time new lattice estimates of the quark mass ratio $z$ appeared and improved the previous ones, namely $z=0.485(20)$ \cite{Fodor:2016bgu}
with three dynamical quarks and $z=0.513(31)$ \cite{Giusti:2017dmp} and
$z=0.453(16)$ \cite{Basak:2018yzz}, with four dynamical quarks.
 By combining them with the older $z=0.470(56)$ \cite{Carrasco:2014cwa} (also with four dynamical quarks),
 we get the following improved estimate 
\begin{equation} \label{eq:znow}
z=\left(\frac{m_u}{m_d}\right)^{\!\!\overline {\rm MS}}\hspace{-10pt}(2~{\rm GeV})=0.472(11)\,,
\end{equation}
which agrees with the previous estimate ($z=0.48(3)$) used in ref.~\cite{diCortona:2015ldu} 
and improves its precision by a factor of three.
We should warn the reader that here the error has been computed by simply propagating the uncertainties quoted by each collaboration, since a proper combination is not yet available. In the remainder of the paper we will use the value in eq.~(\ref{eq:znow}) as reference, however we will always report separately the uncertainty originating from $z$ and the total one so that it can easily be rescaled if needed.

For the LECs appearing in eq.~(\ref{eq:chiNLO}) we now use the values 
\begin{equation}\label{eq:htl7}
h_1^r-h_3^r-\ell_4^r=-0.0049(12)\,,\qquad \ell_7=0.0065(38)\,.
\end{equation}
The first combination is computed using the matching to $L_8^r$ as described 
in ref.~\cite{diCortona:2015ldu}
and using the latest FLAG estimate $L_8^r=0.00055(15)$ \cite{Aoki:2016frl}, 
while the second value is taken 
from the direct lattice simulation of ref.~\cite{Boyle:2015exm}.
These values give
\begin{equation} \label{eq:delta1}
\delta_1=-0.042(13)\,,
\end{equation}
where the error is dominated by the one from $\ell_7$.
Combining everything together we get the updated values for the topological susceptibility and the axion mass at NLO
\begin{equation} \label{eq:chiNLOupdate}
\chi_{\rm top}^{1/4}= 75.46(29)~{\rm MeV}\,,\qquad m_a=5.69(2)_z(4)_{\ell_i^r}\,\mu{\rm eV}\ \frac{10^{12}~{\rm GeV}}{f_a}\,.
\end{equation}
Given the improved value for the light quark mass ratio, the dominant error now became the
one from the NLO LECs, in particular $\ell_7$, which also controls the strong 
isospin breaking effect in the pion mass splitting, indeed poorly known. An improvement on this quantity would directly translate into an equivalent improvement in
our knowledge of $\chi_{\rm top}$ and thus $m_a$. Conversely, improvements in the direct 
computation of $\chi_{\rm top}$ on the lattice could be used to better determine 
both $z$ and $\ell_7$.

A natural question to ask is how much an advance in our knowledge of the light quark masses and the NLO LECs can increase the precision of $\chi_{\rm top}$, before other unknown
corrections need to be considered. Among the latter, the most relevant are the NNLO corrections
of the chiral expansion and ${\cal O}(\alpha_{\rm em})$ electromagnetic (EM) corrections.
The firsts do not only determine the ultimate precision reachable with eq.~(\ref{eq:chiNLO}) but also measure the convergence and reliability of the chiral expansion. Of course the size of the NNLO corrections is only relevant in the chiral expansion approach and does not represent a source of uncertainty for lattice simulations\footnote{On the other hand, lattice simulations
have to face a number of systematic uncertainties which are not present in the chiral expansion
such as finite volume, finite lattice spacing effects, explicit chiral symmetry breaking, etc.,
some of which require delicate and careful analyses.}, which contain the full non-perturbative result. The EM corrections, on the other hand, are common to both approaches and
so far have never been considered. As we will show in the next section, their size is smaller 
with the choice made in eq.~(\ref{eq:chiLO}) of using the value of the neutral pion mass 
in the LO formula.  
Even with this choice, however, the value of the EM corrections is just below 
the size of the present uncertainties for $\chi_{\rm top}$, which means that further improvements cannot ignore them.

In the rest of the paper we will present first the analysis of the EM corrections to the topological susceptibility in sec.~\ref{sec:EMcorr}, and then the NNLO ones in sec.~\ref{sec:NNLOcorr}. We will combine all of them in sec.~\ref{sec:final} where we will also 
give our final estimate for the axion mass with a discussion of the various sources of uncertainties. In appendix~\ref{app:bareresults} we report the formulas for our results with the explicit 
quark mass dependence, suitable to be used in lattice simulation fits. Finally, in appendix~\ref{app:capp} we give all the details of the numerical extraction of the values of the LECs
used in the results.

\section{QED corrections}\label{sec:EMcorr}
While the QCD axion has a vanishing electric charge, its mass can receive ${\cal O}(\alpha_{\rm em})$
corrections from several sources. Indeed, the leading order formula~(\ref{eq:chiLO}) involves a number of quantities that can introduce potentially large EM corrections depending on the way they are defined and extracted by experiments. 
\begin{itemize}
\item
The pion masses for the neutral and the charged states are degenerate at leading order,
but differ at higher orders due to isospin and EM effects. The latter largely dominate this difference, which amounts to $m_{\pi^+}-m_{\pi^0}=4.5936(5)$ MeV, i.e. around 4\% of the total mass. 
 The main effect comes from the charged pion mass, whose corrections are ${\cal O}(e^2)$, while those in the neutral pion mass start at ${\cal O}(e^2 p^2)$. Therefore, depending on which pion mass is used in eq.~(\ref{eq:chiLO}), the axion mass can vary by 4\%, which is more than the quoted 
 uncertainties of the previous section. As we will see below, the naive expectation that the neutral pion mass should be used to minimize EM effects is the correct one. Indeed the pion mass
entering in the leading order formula can be understood as arising from the mixing between the axion and the neutral pion state.

\item 
In QCD the pion decay constant $f_\pi$ is not unambiguously defined when EM interactions are
turned on. In chiral perturbation theory, on the other hand, $\alpha_{\rm em}$ can be
controlled analytically and it is possible to define $f_\pi$ unambiguously. The best determination of $f_\pi$ at the moment comes from (radiative) leptonic pion decays $\pi^+\to \mu{\nu_\mu}(\gamma)$ where both experimental and theoretical uncertainties are small~\cite{Tanabashi:2018oca}. As we will discuss in
more detail below, the EM corrections to $\Gamma_{\pi^+\to\mu\nu(\gamma)}$ are dominated by a 
calculable short distance contribution. The long distance hadronic contribution (which is 
of the same order of the EM corrections we want to compute for the axion mass) is subleading
but dominates the current error of $f_\pi$. Given the importance of such corrections for our
computation, we revisit their estimate and analyze their interplay with the genuine corrections
to the axion mass. An alternative determination of $f_\pi$ could be obtained from the neutral pion decay $\pi^0\to\gamma\gamma$, however both the theoretical and experimental uncertainties are not competitive with the charged pion channel~\cite{Tanabashi:2018oca}.
 
\item
While the light quark mass ratio $z=m_u/m_d$ at leading order is renormalization group (RG) 
invariant with respect to QCD corrections, it is not with respect to the 
QED ones~\cite{Gasser:1982ap}. 
This introduces an ${\cal O}(\alpha_{\rm em})$ ambiguity in the tree-level formula of the axion
mass that should be removed by the sub-leading EM corrections:
\begin{equation}
\frac{\partial\log z}{\partial \log \mu} =\frac{6\alpha_{\rm em}}{4\pi}
\left[\left (\frac23\right )^2-\left( -\frac13 \right)^2 \right] =
\frac{\alpha_{\rm em}}{2\pi}
\end{equation}
A change of ${\cal O}(1)$ in the renormalization scale introduces a shift of ${\cal O}(10^{-3})$
in $z$ that can be taken as a lower bound to the order of magnitude of the expected EM corrections to the axion mass.
\end{itemize}

We start by reporting the result for the computation of the leading EM corrections to the topological susceptibility, which begin at ${\cal O}(e^2 p^2)$ in the chiral expansion 
once the leading order term is written in terms of the physical\footnote{Whenever $m_\pi$ appears in the following formulas, it can be equivalently understood as $m_{\pi^0}$ or $m_{\pi^+}$ because the difference will be accounted by higher orders in either $e^2$ or $p^2$. For the numerical estimates we used $m_{\pi}=m_{\pi^0}$.} neutral pion mass $m_{\pi^0}$ (including EM corrections) and the physical charged pion decay constant $f_{\pi^+}$ (defined in pure QCD, i.e. at  $\alpha_{\rm em}=0$):
\begin{align} \label{eq:emcorr}
\chi_{\rm top}&=\frac{z}{(1+z)^2}\,
m_{\pi^0}^2 f_{\pi^+}^2 \left[ 1+\delta_e + \dots \right ]\,, \\
\delta_e&=e^2 \left[\frac{20}{9}(k_1^r+k_2^r)-4k_3^r+2k_4^r+\frac{8}{3}\Delta\, k_7^r-\frac{Z}{4\pi^2}\left(1+\log\left(\frac{m_{\pi}^2}{\bar\mu^2}\right)\right) \right]\,, \label{eq:deltae}
\end{align}
where dots in eq.~(\ref{eq:emcorr}) represent the non-EM corrections discussed in secs.~\ref{sec:LONLO} and~\ref{sec:NNLOcorr}, the coefficients $Z$ and $k_i^r$ are the $n_f=2$ EM low-energy constants from~\cite{Knecht:1997jw},
and $\bar \mu$ is the renormalization scale of the chiral Lagrangian, whose dependence cancels against that
from the $k^r_i$ coefficients. 
As anticipated before, once the LO formula is written in terms of $m_{\pi^0}^2$, the
EM corrections start at ${\cal O}(e^2 p^2)$ (the $\delta_e$ term).
In particular the EM pion mass splitting effects parametrized by
\begin{equation}
Z=\frac{m_{\pi^+}^2-m_{\pi^0}^2}{2e^2 f_{\pi^+}^2}+\dots \simeq 0.81\,.
\end{equation}
are loop suppressed.
Although the value for the couplings $k_i^r$ is not known directly, it can been inferred, as in~\cite{Haefeli:2007ey}, using their relation to the $n_f=3$ constants $K_i^r$, which have been estimated in refs.~\cite{Ananthanarayan:2004qk,Bijnens:1996kk} using various techniques including sum rules and vector meson dominance.
The values for the $k_i^r$ we use are taken from~\cite{Haefeli:2007ey} (with $K_9^r$ from~\cite{Bijnens:1996kk}) and reported in tab.~\ref{tab:ks}.
\begin{table}[t]
\centering
\begin{tabular}{ c c c c c | c }
\hline 
$ k_1^r$ & $ k_2^r$ & $ k_3^r$ & $ k_4^r$ & $ k_7^r$ & \\ \hline
$ 8.4$   & $ 3.4$   & $ 2.7$   & $ 1.4$   &   $2.2$ & $\times 10^{-3}$\\ \hline
\end{tabular}
\caption{\label{tab:ks} \emph{Numerical values of the $n_f=2$ EM LECs $k_i^r$ at the scale $\bar \mu=770$~MeV extracted using their relation to $K_i^r$. To the $k_i^r$ it is assigned a conservative $100\%$ uncertainty.}}
\end{table}
Because of the model dependence of such estimates we decided to assign a conservative 100\% uncertainty to each LEC, i.e. we use the mentioned values as an order of magnitude estimate
of their size. Substituting the numerical values we find
\begin{equation}\label{eq:deltaenum}
\delta_{e}=0.0065(21)\,.
\end{equation}
While we have assigned 100\% uncertainties to the LECs $k_i^r$, the uncertainty on $\delta_e$
only amounts to 30\% because the dominant contribution comes from the last term  
in eq.~(\ref{eq:deltae}).

As discussed before, the QED RG scale dependence from the quark mass ratio $z$  in the leading
order formula~(\ref{eq:chiLO}) must be reabsorbed by ${\cal O}(\alpha_{\rm em})$ corrections. 
Indeed the EM LECs $k_{5,7}^r$ have the non-trivial UV-scale $\mu$ dependence\footnote{This can 
be derived by computing the operators generated in the chiral Lagrangian 
by an RG transformation of the quark mass matrix in terms of the EM charge spurions
$Q_{L,R}$.}:
\begin{equation} \label{eq:k5k7}
\mu\frac{\partial}{\partial \mu} k_5^r =  -\frac35 \frac{1}{(4\pi)^2}  \,,\qquad 
\mu\frac{\partial}{\partial \mu} k_7^r =  -\frac34 \frac{1}{(4\pi)^2}  \,.
\end{equation}
It is easy to check that the variation of $k^r_7$ reabsorbs the dependence induced 
by the variation of $z$ in the leading order formula (in the $n_f=3$ case the RG scale 
dependence is reabsorbed by $K_9^r$). In fact, the light quark mass ratio $z$ 
and the constants $k^r_{5,7}$ cannot be determined independently and only the RG invariant
combination enters physical quantities. The numerical value of $k_7^r$ in tab.~\ref{tab:ks} 
is of the same order of the scale dependence in eq.~(\ref{eq:k5k7}), which therefore dominates
its determination. In any case, the current uncertainties on the quark mass
ratio $z$ are still bigger than the effects from the scale dependence in $z$, and therefore bigger than the effects from $k^r_{7}$.

To complete the computation of $\chi_{\rm top}$ we need the value of the pion decay constant
$f_{\pi^+}$ at $\alpha_{\rm em}=0$. Currently the best determination comes from the 
charged pion leptonic decay, which according to the PDG~\cite{Tanabashi:2018oca} provides $f_{\pi^+}=92.28(9)$. 
This estimate however involves EM corrections of the same order of $\delta_e$, so that
a consistent calculation of $\chi_{\rm top}$ within the chiral expansion should consider the
two sources of EM corrections together. In more details $f_{\pi^+}$ is related to the EM
inclusive pion decay rate via
\begin{equation}\label{eq:Gammapi}
 \Gamma_{\pi^+\to \mu \nu(\gamma)}=\frac{G_F^2 |V_{ud}|^2  m_{\pi^+} m_\mu^2 f_{\pi^+}^2}{4\pi}
\Bigl( 1-\frac{m_\mu^2}{m_{\pi^+}^2}\Bigr)^2
\, \left[1+\delta_\Gamma^{\rm loc}+\delta_\Gamma^{\rm had} \right]
 \end{equation} 
where the $\delta_\Gamma$ terms computed in~\cite{Cirigliano:2007ga} are the ${\cal O}(\alpha_{\rm em})$
corrections, which we split into two terms: the local contribution $\delta_\Gamma^{\rm loc}$
and the IR one $\delta_\Gamma^{\rm had}$, which parametrizes the hadronic form factors 
and depends on the chiral LECs.
Explicitly they read:
\begin{align}
\delta_\Gamma^{\rm loc} &  =  \frac{\alpha_{\rm em}}{\pi} \left[
\log\left( \frac{m_Z^2}{m_\rho^2} \right)+F\left(\frac{m_\mu^2}{m_{\pi^+}^2} \right)
-\frac{m_\mu^2}{m_\rho^2} \left( {\bar c}_2 \log\left( \frac{m_\rho^2}{m_\mu^2}\right)+{\bar c}_3+{\bar c}_4\right) + \frac{m_{\pi^+}^2}{m_\rho^2} {\bar c}_{2t} \log \left( \frac{m_\rho^2}{m_\mu^2} \right)
\right] , \nonumber \\
\delta_\Gamma^{\rm had} & =  e^2 \left[ \frac83(K_1^r+K_2^r)+\frac{20}{9} (K_5^r+K_6^r)
-\frac43 X_1^r-4(X_2^r-X_3^r)-\tilde X_6^{r,\rm eff} \right. \nonumber \\ 
& \left. \qquad + \frac{1}{(4\pi)^2}\left(
2-3Z-Z\log\left(\frac{m_K^2}{\bar \mu^2} \right) + (3-2Z)\log\left(\frac{m_\pi^2}{\bar\mu^2}\right)
\right)\right] , \nonumber \\
F(x)&\equiv \frac32 \log x+\frac{13-19x}{8(1-x)}-\frac{8-5x}{4(1-x)^2}x\log x
-\left(2+\frac{1+x}{1-x}\log x\right) \log(1-x)-2\frac{1+x}{1-x} {\rm Li}_2(1-x)\, .
\end{align}  
The constants $\bar c_i$ have been estimated in a model dependent way 
in ref.~\cite{Cirigliano:2007ga}, and for this reason we assign a conservative 
100\% uncertainty to them. The corresponding numerical values
from~\cite{Cirigliano:2007ga} are reported in tab.~\ref{tab:cbarKXs}.
\begin{table}[t]
\centering
\begin{tabular}{ c c c c }
\hline 
$\bar{c}_2$ & $\bar{c}_3$ & $\bar{c}_4$ & $\bar{c}_{2t}$ \\ \hline
$ 5.2$ & $ -10.5$ &$ 1.69$ &$0$\\ \hline
\end{tabular} \\[12pt]
\begin{tabular}{ c c c c c c c c | c ccc|c}
\hline 
$ K_1^r$ & $  K_2^r$ & $  K_3^r$ & $  K_4^r$ & $  K_5^r$ & $  K_6^r$ & $  K_9^r$ & $  K_{10}^r$  &
$  X_1^r$ & $  X_2^r$ & $  X_3^r$ & $  \tilde X_6^{r,\rm eff}$
\\ \hline
$ -2.7$ & $ 0.7$ & $ 2.7$ & $ 1.4$ & $ 12$ & $ 2.8$ & $ -1.3$ & $ 4$ & 
$ -3.7$ & $ 3.6$ & $ 5$ & $ 13$ &
$\times 10^{-3}$ \\ \hline
\end{tabular}
\caption{\label{tab:cbarKXs} \emph{Top: Numerical values of the $\bar{c}_i$ constants appearing in $\delta_\Gamma^{\rm loc}$ from~\cite{Cirigliano:2007ga}. 
Bottom: Numerical values of the $n_f=3$ radiative and leptonic LECs from~\cite{Ananthanarayan:2004qk,Bijnens:1996kk,DescotesGenon:2005pw} at the scale $\bar \mu=770$~MeV.
All constants are assigned a conservative $100\%$ uncertainty.}}
\end{table}
The constants $K_i^r$ and $X_i^r$ are the $n_f=3$ radiative and leptonic LECs respectively,
defined in refs.~\cite{Urech:1994hd,Knecht:1999ag} (except for $\tilde X_6^{r,{\rm eff}}$ defined in ref.~\cite{DescotesGenon:2005pw}). We use the values estimated in ref.~\cite{Ananthanarayan:2004qk} for $K_{1,\dots,6}^r$, in ref.~\cite{Bijnens:1996kk} for $K_{9,10}^r$, and in ref.~\cite{DescotesGenon:2005pw} for $X_i^r$, which we report in tab.~\ref{tab:cbarKXs} and to which we associate conservatively a 100\% uncertainty.\footnote{The various estimates of the $K_i^r$ in
refs.~\cite{Ananthanarayan:2004qk,Bijnens:1996kk} and references therein are not always compatible
with each other, hence our conservative choice for the error, which is supposed to take those
model-dependent deviations into account.}

Numerically the size of the EM corrections to $\Gamma_{\pi^+\to \mu \nu(\gamma)}$ amounts to
\begin{equation} \label{eq:deltaGchi}
\delta_\Gamma^{\rm loc}+\delta_\Gamma^{\rm had} = 0.0177(38)\,,
\end{equation}
very close to the PDG estimate (0.0176(21)) but with larger error (here we have been more conservative). Note that, although the uncertainties of all
LECs have been taken ${\cal O}(1)$, the result has a 20\% accuracy, since the first term 
in $\delta_\Gamma^{\rm loc}$ largely dominates over all the others. Combining eqs.~(\ref{eq:Gammapi}) and~(\ref{eq:deltaGchi})
we get $f_{\pi^+}=92.26(18)$. 

Since some of the LECs appearing in $\delta_\Gamma^{\rm had}$ are common with some of those
appearing in $\delta_e$, the topological susceptibility $\chi_{\rm top}$ should be written
directly in terms of the  $\Gamma_{\pi^+\to \mu \nu(\gamma)}$ rather than $f_{\pi^+}$, i.e.
\begin{equation}\label{eq:chiEMonly}
\chi_{\rm top}=\frac{z}{(1+z)^2}m_{\pi^0}^2 
\frac{ \Gamma_{\pi^+\to \mu \nu(\gamma)}}{\frac{G_F^2 |V_{ud}|^2  m_{\pi^+} m_\mu^2}{4\pi}\Bigl( 1-\frac{m_\mu^2}{m_{\pi^+}^2}\Bigr)^2}\left[1+\delta_e-\delta_\Gamma^{\rm loc}-\delta_\Gamma^{\rm had}+\dots\right] \, .
\end{equation}
The combination $\delta_e-\delta_\Gamma^{\rm had}$ can be written either all entirely in terms 
of $n_f=3$ LECs, or in a hybrid way in terms of $n_f=2$ $k_i^r$ and $n_f=3$ $X_i^r$ 
(because the $n_f=2$ leptonic LECs are not available):
\begin{align}
\delta_e-\delta_\Gamma^{\rm had} &=
e^2 \left[2k_4^r-4k_3^r+\frac{8}{3}\Delta\, k_7^r-4k_9^r  
+\frac43 X_1^r+4(X_2^r-X_3^r)+\tilde X_6^{r,\rm eff} 
- \frac{
2(1+Z) + (3+2Z)\log \frac{m_\pi^2}{\bar\mu^2} }{(4\pi)^2}\right] \nonumber \\
&=e^2 \left[ 2K_4^r-4K^r_3 +\frac83\Delta(K_9^r+K_{10}^r)+\frac43 X_1^r+4(X_2^r-X_3^r)
+\tilde X_6^{r,\rm eff} \right. \nonumber \\
&\qquad \left. -\frac{1}{(4\pi)^2} \left ( 
2(1+Z+\Delta Z)+(3+2Z)\log\frac{m_\pi^2}{\bar \mu^2}+2\Delta Z \log\frac{m_K^2}{\bar\mu^2}
\right)
 \right]\,.
\end{align}
In this way we get
\begin{equation}\label{eq:deltadeltanum}
\delta_e-\delta_\Gamma^{\rm had}-\delta_\Gamma^{\rm loc}=0.024(6)\,,
\end{equation}
which in combination with eq.~(\ref{eq:chiEMonly}) can be used to evaluate the EM contribution to $\chi_{\rm top}$.

The direct extraction of $f_{\pi^+}$ from lattice QCD simulations is not competitive with the estimate above. However, recently the EM corrections to $\Gamma_{\pi^+\to \mu \nu(\gamma)}$ have been computed in a preliminary study on the lattice~\cite{Lubicz:2016mpj}, giving\footnote{Note however that ref.~\cite{Giusti:2017dwk} alerts about upcoming results which slightly deviate from this quoted value.}
\begin{equation} \label{eq:deltaGlat}
\delta_\Gamma^{\rm loc}+\delta_\Gamma^{\rm had} = 0.0169(15)\,,
\end{equation}
which is in very good agreement with eq.~(\ref{eq:deltaGchi}). Accidentally this value is 
very close to the PDG one, both in size and uncertainty.
Eq.~(\ref{eq:deltaGlat}) implies $f_{\pi^+}=92.30(7)$. Given the compatibility of the chiral
and the lattice results, and the fact that the latter has better precision and 
less model dependence, we will use eq.~(\ref{eq:deltaenum}) and this lattice estimation for $f_{\pi^+}$, bearing in mind that
numerically this choice is also equivalent to using the PDG determination.

\section{NNLO corrections}\label{sec:NNLOcorr}
Given the smallness of the expansion parameter, the $n_f=2$ chiral expansion is expected to converge very fast and the NNLO corrections to be only a few percent with respect to the NLO ones. Nevertheless, depending on the magnitude of the low-energy coefficients, they might be competitive with the EM corrections discussed in the previous section. Their estimation is therefore essential for a precise calculation of $\chi_{\rm top}$.

\begin{figure}
\centering
\includegraphics[scale=0.75]{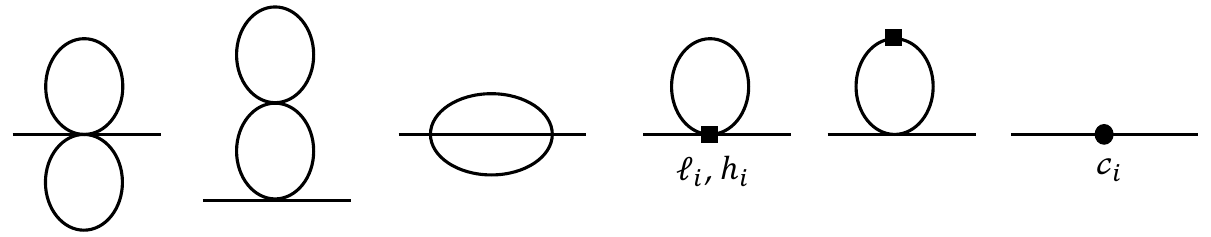}
\caption{\label{fig:diag} \emph{One-particle-irreducible diagrams for the axion and pion $2$-point functions at~NNLO.}}
\end{figure}

The NNLO corrections to the topological susceptibility receive contributions from the diagrams in fig.~\ref{fig:diag}, which correspond to: 1) the two-loop diagrams constructed from $\mathcal{O}(p^2)$ vertices, 2) the one-loop diagrams generated by $\mathcal{O}(p^4)$ vertices and 3) the tree-level graphs from the $\mathcal{O}(p^6)$ Lagrangian.\footnote{Note that even with a quark field redefinition that avoids the tree-level mixing between the axion and the neutral pion, this mixing arises at one loop, producing effects at NNLO in $\chi_{\rm top}$.} 
At $\alpha_{\rm em}=0$, the full NNLO result is
\begin{align}\label{eq:NNLOcorr}
\chi_{\rm top}^{\rm NNLO}&=\frac{z}{(1+z)^2}\,
m_{\pi^0}^2 f_{\pi^+}^2 \left[ 1+\delta_1 +\delta_2 \right ]\,, \\
\delta_2 & =  \frac{m_{\pi}^4}{f_{\pi}^4} \left[ 32(c_6^r+2c_{19}^r) +\frac{2  \bar{\ell}_3+4  \bar{\ell}_4+3 \log\frac{m_\pi^2}{\bar\mu^2}-2}{4(4\pi)^4}  \log\frac{m_\pi^2}{\bar\mu^2}  +\frac{3\ell_7}{8\pi^2} \log\frac{m_\pi^2}{\bar\mu^2}\right. \nonumber \\ 
&\left. \qquad \qquad + \frac{h_1^r-h_3^r-\ell_4^r-\ell_7}{8\pi^2}\bar{\ell}_3 
-\frac{1}{(4\pi)^4}\left(\bar{\ell}_4^2-\bar{\ell}_3 \bar{\ell}_4+\frac{25}{16}\right)\right. \nonumber \\
&\left. \qquad \qquad +\Biggl(32 \left(c_9^r-6 c_{10}^r-4 \left(c_{11}^r+c_{17}^r+c_{18}^r\right)-6 c_{19}^r\right)-\frac{7\ell_7}{8\pi^2}\log\frac{m_\pi^2}{\bar\mu^2}\right. \nonumber \\
& \left. \qquad \qquad + 8(h_1^r-h_3^r-\ell_4^r)\ell_7 -12 \ell_7^2+\frac{2 \bar{\ell}_3+2 \bar{\ell}_4-1 }{8\pi^2}\ell_7\Biggr)\Delta^2+20\ell_7^2\, \Delta^4\right]\,,  \label{eq:delta2}
\end{align}
with $\delta_1$ given in eq.~(\ref{eq:chiNLO}) and the $c_i^r$ being the $\mathcal{O}(p^6)$ $n_f=2$ LECs introduced in~\cite{Bijnens:1999sh,Bijnens:1999hw}. Note that the charged and neutral pion decay constants (defined, as mentioned, at $\alpha_{\rm em}=0$) differ only at two loops due to isospin breaking effects, so $f_\pi$ in $\delta_1$ and $\delta_2$ can be understood either as $f_{\pi^+}$ or $f_{\pi^0}$, being the difference accounted by $\mathcal{O}(p^8)$ terms. The scale dependence of the combinations of $c_i^r$ in eq.~(\ref{eq:delta2}) is fully reabsorbed by the $\log\frac{m_\pi^2}{\bar\mu^2}$ and $\log^2\frac{m_\pi^2}{\bar\mu^2}$ terms, and since $h_1^r-h_3^r-\ell_4^r$, $\ell_7$ and $\bar{\ell}_{3,4}$ are scale invariant, the scale dependence of $\delta_2$ cancels separately in each line of~eq.~(\ref{eq:delta2}).

While the numerical value of most of the $\mathcal{O}(p^4)$ LECs is reasonably well known, the determination of the $c_i^r$ is in much worse shape. In fact, only few combinations of $c_i^r$ can be extracted directly because there are not enough experimental observables to fit all the $n_f=2$ Lagrangian parameters.\footnote{In fact, of the $c_i^r$ that appear in eq.~(\ref{eq:delta2}), only $c_6^r$ has been extracted semi-directly from experiments, in particular from the pion scalar form factor~\cite{Bijnens:1998fm}, with some phenomenological modeling. As explained in app.~\ref{app:capp}, since on its numerical value there is still disagreement~\cite{Donoghue:1990xh,Bijnens:1998fm,Caprini:2018jjm}, we will not use it in our analysis.} Recent partially-quenched lattice QCD simulations~\cite{Boyle:2015exm} provided results 
for some of the combinations of $c_i^r$ appearing in eq.~(\ref{eq:delta2}). For the remaining ones
we matched the relevant combinations to the $n_f=3$ LECs, for which some estimates exist~\cite{Bijnens:2014lea} (taken with a conservative $100\%$ error). In this way we have been able 
to extract  an order of magnitude estimate for all the $c_i^r$ appearing in eq.~(\ref{eq:delta2}), which we report in tab.~\ref{tab:cs} (see app.~\ref{app:capp} for more details).

\begin{table}[t]

\centering
\begin{tabular}{ccccccccc}
\hline 
$ 10^6c_6^r$ & $ 10^6c_7^r$ & $ 10^6c_8^r$ & $ 10^6c_9^r$ & $ 10^6c_{10}^r$ & $ 10^6c_{11}^r$ & $ 10^6c_{17}^r$ & $ 10^6c_{18}^r$ &$ 10^6c_{19}^r$ \\ \hline
$ 1.0(3.8)$ & $ -2.2(2.0)$ & $ 2.1(1.1)$ & $ -0.3(1.2)$ & $ -1.0(1.1)$ & $ -0.1(0.7)$ & $ 6.4(3.8)$ & $ -2.2(5.9)$ &$ -0.1(9.5)$ \\ \hline
\end{tabular}
\caption{\label{tab:cs} \emph{Numerical values  of the $n_f=2$ LECs in $\delta_2$ at the scale $\bar \mu=770$~MeV  extracted by combining the lattice results of~\cite{Boyle:2015exm} and the matching with the $n_f=3$ LECs of~\cite{Bijnens:2014lea} (see app.~\ref{app:capp} for more details).}}
\end{table}

The LECs in tab.~\ref{tab:cs} and eq.~(\ref{eq:htl7}), combined with the values $\bar{\ell}_3=2.81(49)$ and $\bar{\ell}_4=4.02(25)$ from~\cite{Boyle:2015exm}, lead to the following numerical result for the NNLO corrections:
\begin{equation} \label{eq:delta2num}
\delta_{\rm 2}= -0.0071(01)_z(23)_{\ell_i^r}(19)_{c_i^r}=-0.0071(29)\,  .
\end{equation}
While the uncertainty from $z$ is very small, those from $\ell_i^r$ (of which $\ell_7$ provides the largest contribution) and $c_i^r$ have similar size. Notice that although the relative uncertainties of the $c_i^r$ are large, they only have a milder impact on the final uncertainty of $\delta_2$, because numerically $\delta_2$ receives bigger contributions from the ${\cal O}(p^4)$ LECs and the
non-local contributions. 
 Moreover, the isospin-breaking terms in $\delta_2$ (the last two lines in eq.~(\ref{eq:delta2})),
 which are suppressed by powers of $\Delta^2\approx 0.1$,
  contribute less than 20\% to the final result and are within the uncertainty of $\delta_2$. As a consequence, the precision on the LECs is still not enough for the result to be sensitive 
  to isospin breaking corrections.
Finally notice that $\delta_2$ is numerically of the same order of the EM corrections in eq.~(\ref{eq:deltaenum}), but with opposite sign. Therefore, both have to be considered for a sub-percent estimate of $\chi_{\rm top}$.

\section{Final Results and Axion Mass}\label{sec:final}
We can now combine the analysis of secs.~\ref{sec:LONLO}, \ref{sec:EMcorr} and \ref{sec:NNLOcorr} and estimate of the topological susceptibility to $\mathcal{O}(p^6,e^2p^2)$. The final result reads
\begin{equation} \label{eq:chitop}
\chi_{\rm top}=\frac{z}{(1+z)^2}\,
m_{\pi^0}^2 f_{\pi^+}^2 \left[ 1+\delta_1 +\delta_2+\delta_e \right ]\,, 
\end{equation}
where the $\mathcal{O}(p^4)$ contribution $\delta_1$ is given in eq.~(\ref{eq:chiNLO}), the $\mathcal{O}(p^6)$ contribution $\delta_2$ in eq.~(\ref{eq:delta2}) and  the $\mathcal{O}(e^2p^2)$ contribution $\delta_e$ in eq.~(\ref{eq:deltae}). For completeness, in app.~\ref{app:bareresults} we also report $\chi_{\rm top}$ expressed in terms of quark masses and bare chiral Lagrangian parameters.

Substituting our numerical estimates, the final results for the topological susceptibility and
the axion mass read
\begin{equation}
\chi_{\rm top}^{1/4}=75.44(34)~{\rm MeV}\,, \qquad 
m_a=5.691(51)\, \mu{\rm eV}\, \frac{10^{12}~{\rm GeV}}{f_a}\,.
\end{equation}
Notice how these values almost coincide with the updated NLO ones in eq.~(\ref{eq:chiNLOupdate}), 
since both NNLO and EM corrections are comparable but smaller than the present uncertainties 
of the NLO estimate, and, having opposite sign, they tend to cancel each other. This
result confirms the reliability of the NLO estimate in \cite{diCortona:2015ldu}.
It is instructive to deconstruct the contributions at each order with the various 
uncertainties: for the axion mass case they read
\begin{equation}
m_a=\Bigl[\underbrace{5.815(22)_z(04)_{f_\pi}}_\text{LO}
\underbrace{-0.121(38)_{\ell^r_i}}_\text{NLO}
\underbrace{-0.022(07)_{\ell^r_i}(05)_{c_i^r}}_\text{NNLO}
\underbrace{+0.019(06)_{k_i^r}}_\text{EM}\Bigr] 
\, \mu{\rm eV}\, \frac{10^{12}~{\rm GeV}}{f_a}\,,
\end{equation}
where the reported uncertainties on each contributions come from those of $z$, 
from the EM corrections in the extraction of $f_{\pi^+}$, and from those of the various LECs 
in the NLO ($\ell^r_i$ and $h_i^r$), in the NNLO ($c_i^r$) and in the EM ($k_i^r$) chiral Lagrangians.

Several comments are in order. First notice how, while NLO corrections are almost two
orders of magnitude smaller than the LO result, NNLO are barely one order of magnitude
below the NLO ones. On one side this means that the chiral expansion is nicely converging
and, given the current uncertainties on $z$ and the LECs, the NLO result is enough. 
On the other side, the size of the NNLO corrections is such that they cannot be ignored 
in future improvements of $m_a$.

EM corrections are of similar size, slightly less than 0.5\% and within present uncertainties.
The numerical estimate of the EM corrections has been carried out using the lattice QCD results for
$f_{\pi^+}$ extracted from eqs.~(\ref{eq:Gammapi}) and~(\ref{eq:deltaGlat}) with $\delta_e$ in eq.~(\ref{eq:deltaenum}), since these values are more model-independent.\footnote{A very similar
result would follow using the PDG value for $f_{\pi^+}$, which, as mentioned, 
is very close to the lattice estimate both in size and error.} However, one
could have also used eqs.~(\ref{eq:chiEMonly}) and (\ref{eq:deltadeltanum}) obtaining essentially the same central value although with an error twice as large.
As for NNLO corrections, they must be considered should the uncertainties coming from 
$z$ and the NLO LECs decrease. As commented before, the size of these corrections also 
represents the ultimate precision that can be reached in lattice estimates which do not include
EM corrections.

We conclude by noticing that, if the uncertainties in $m_u/m_d$ and the NLO LECs (in particular $\ell_7$) are reduced by a factor of few (which is not unreasonable) our results could be
used to determine the axion mass (and $\chi_{\rm top}^{1/4}$) 
with per-mille accuracy.

\section*{Acknowledgements}

We are grateful to Johan Bijnens for correspondence and clarifications on the formulas for the pion mass and the decay constant in refs.~\cite{Ananthanarayan:2017yhz,Amoros:1999dp}. We thank Gilberto Colangelo for pointing out a typo in Eq.~(32) of App.~A in the previous version of the manuscript.

\appendix


\section{Results in terms of the quark masses}\label{app:bareresults}
We provide here the results for pion mass, pion decay constant and topological susceptibility in both $n_f=2$ and $n_f=3$ chiral perturbation theory (in this last case in the unbroken isospin limit $m_u=m_d$) in terms of the bare chiral Lagrangian parameters and quark masses. These are intermediate results used to obtain the formulas in the main sections and the matching of app.~\ref{app:capp}.

\subsection{Two-flavor results}
We start with the neutral pion mass $m_{\pi^0}$, calculated at $\mathcal{O}(p^6,p^2e^2)$, and the charged pion decay constant $f_{\pi^+}$, defined at $\alpha_{\rm em}=0$ and calculated at $\mathcal{O}(p^6)$. These have been calculated in the unbroken isospin limit in~\cite{Burgi:1996qi,Bijnens:1997vq,Colangelo:2001df,Bijnens:2017wba,Knecht:1997jw}, and read:
\begin{align}\label{eq:mpibare}
m_{\pi^0}^2 = & \  M^2  \left[1 + \delta_1^M+\delta_2^M+\delta_e^M\right]  \, ,\\
f_{\pi^+} = &  \ F  \left[1 +\delta_1^F+\delta_2^F\right] \, ,\label{eq:fpibare}
\end{align}
where $M^2\equiv B (m_u+m_d)$, $B$ and $F$ are the LECs of the leading order $n_f=2$ Lagrangian~\cite{Gasser:1983yg}, and
\begin{align}
\delta_1^M =&\ \frac{M^2}{F^2}\left[ (2\ell_3^r +(1/2)\kappa \lambda_M ) -   (2\ell_7) \Delta^2\right]	  \, ,\\
\delta_1^F =&\ \frac{M^2}{F^2} \left[ \ell_4^r -  \kappa\, \lambda_M  \right] \, , \\
\delta_e^M = &\ e^2\left[- \frac{20}{9}	\left(k_1^r+k_2^r-\frac{9}{10}(2k_3^r-k_4^r)-k_5^r-k_6^r-\frac15 (1-3\Delta)k_7^r\right) + 2\kappa Z\left(1+\lambda_M\right) \right] ,
\end{align}
where $\kappa\equiv(4\pi)^{-2}$ and $\lambda_M\equiv\log\frac{M^2}{\bar{\mu}^2}$. The $\mathcal{O}(p^6)$ contributions are
\begin{align}
\delta_2^M= \frac{M^4}{F^4}\bigg\{&-16\left(2c_6^r+c_7^r+2c_8^r+c_9^r-3c_{10}^r-2\left(3c_{11}^r+c_{17}^r+2c_{18}^r\right)\right)
\\ \nonumber 
 & +  (\ell_1^r+2 \ell_2^r+\ell_3^r)\, \kappa+\frac{163}{96} \kappa^2 +\left[-(14 \ell_1^r+8 \ell_2^r+3 \ell_3^r)\, \kappa-\frac{49}{12}\kappa^2\right]\lambda_M+ \frac{17}{8}\kappa^2 \lambda_M^2  \nonumber\\ 
 & +\Big[
 -16\,\big(c_{7}^{r} + c_{9}^{r} - 9c_{10}^{r} - 6c_{11}^{r} - 6c_{17}^{r} - 4c_{18}^{r} - 8c_{19}^{r}\big)+\kappa(1+5\lambda_M)\ell_7 \Big]\Delta^2\bigg\} \, \nonumber, \\ \nonumber 
\nonumber\\
\delta_2^F=\frac{M^4}{F^4}\bigg\{& 8\left( c_7^r+2 c_8^r+ c_9^r\right)+\left(-(1/2) \, \ell_1^r-\ell_2^r-2 \ell_3^r\right) \, \kappa-\frac{13}{192}\kappa^2 \ \\ \nonumber 
&+\left[(7 \ell_1^r+4 \ell_2^r-2 \ell_3^r-(1/2)\, \ell_4^r)\, \kappa+\frac{23}{12} \kappa^2\right]\lambda_M-\frac{5}{4}\kappa^2\lambda_M^2 \ \\ \nonumber
&+ \Big[8\left(c_7^r-c_9^r\right)+\kappa\,(1+\lambda_M)\ell_7\Big]\Delta ^2\bigg\}\nonumber\, .
\end{align}

At the same order, the topological susceptibility reads 
\begin{align}\label{eq:topbare}
\chi_{\rm top} = & \  \frac{z}{(1+z)^2}M^2 F^2  \left[1 + \delta_1^\chi+\delta_2^\chi+\delta_e^\chi\right]  \, ,
\end{align}
where the NLO correction have been first computed in ref.~\cite{Mao:2009sy}
\begin{align}
\delta_1^\chi =& \ \frac{M^2}{F^2}\left[ 2 \left(h_1^r-h_3^r- \ell_7+\ell_3^r\right)-(3/2)\kappa \lambda_M	+\left(2\ell_7\right)\Delta^2 \right] \, , \end{align}
while the EM and the NNLO corrections read
\begin{align}
\delta_e^\chi =&\ e^2\left[  \frac{20}{9}  \left(k_5^r+k_6^r\right)+\frac49 (1+3\Delta) \, k_7^r -2\kappa Z\left(1+\lambda_M\right) \right] \, ,\\
\delta_2^\chi =&\ \frac{M^4}{F^4}\Big\{16\left (3 c_ {10}^r + 6 c_ {11}^r + 2 c_ {17}^r + 4\left (c_{18}^r + c_ {19}^r \right) \right) - 3\kappa\ell_3^r +3\kappa\left[-(1/4)\kappa-3\ell_3^r+2\ell_7\right]\lambda_M \  \\
    \ &-\frac98\kappa^2\lambda_M^2+\left[-48 c_{10}^r-32 \left(c_{11}^r+c_{17}^r+2 \left(c_{18}^r+c_{19}^r\right)\right)+\kappa \left(1-7 \lambda_M\right)\ell_7^r\right]\Delta^2 \Big\} \, .\nonumber
\end{align}

It is a nontrivial consistency check that the dependence on the scale $\bar{\mu}$ cancels separately in any of the previous equations. Moreover, the QED running of the quark masses is compensated by the shift of $k_i^r$ as explained in section~\ref{sec:EMcorr}, in such a way that both $m_{\pi^0}$ and $\chi_{\rm top}$ are independent of the QED RG scale $\mu$. Inverting eqs.~(\ref{eq:mpibare}-\ref{eq:fpibare}) for $M$ and $F$ and plugging the result into eq.~(\ref{eq:topbare}), we obtain the topological susceptibility $\chi_{\rm top}$ expressed as a function of the physical $\pi^0$ mass and $f_{\pi^+}$ only, as in eq.~(\ref{eq:chitop}).

\subsection{Three-flavor results}\label{subsec:3flavbare}
In the unbroken isospin limit $m_u=m_d\equiv m$ and at $\alpha_{\rm em}=0$, the pion mass and decay constant at NNLO in $n_f=3$ chiral perturbation theory are
\begin{align}\label{eq:mpibare3f}
m_{\pi}^2 = & \  M_0^2  \left[1 + \epsilon_1^M+\epsilon_2^M\right]\, ,\\
f_{\pi} = &  \ F_0  \left[1 +\epsilon_1^F+\epsilon_2^F\right]  \, ,\label{eq:fpibare3f}
\end{align}
where $M_0^2\equiv2 B_0 m$, $B_0$ and $F_0$ are the LECs of the leading order $n_f=3$ Lagrangian of~\cite{Gasser:1984gg} and 
\begin{align}
\epsilon_1^M =&\ -\frac{B_0 m_s}{F_0^2} \left\{2\left[\frac{ \kappa  \lambda _{\eta }}{9}+8\left(L_4^r-2 L_6^r\right)\right]+\left[\kappa  \left(\frac{\lambda _{\eta }}{9}- \lambda _0\right)+16\left(2 L_4^r+ L_5^r-4 L_6^r-2 L_8^r\right)\right]w\right\}\, , \\
\epsilon_1^F =&\ -\frac{B_0 m_s}{F_0^2} \left\{\left[\frac{\kappa  \lambda _K}{2}-8 L_4^r\right]+\left[\kappa\left(2  \lambda _0+\frac{ \lambda _K}{2}\right)-8\left(2 L_4^r+ L_5^r\right)\right]w\right\}
\end{align}

Here $m_s$ is the strange quark mass, $w\equiv m/m_s$, $M_K^2\equiv B_0 (m+m_s)$ and $M_\eta^2\equiv (4/3)M_K^2-(1/3)M_0^2$ are the tree-level kaon and eta masses, $\lambda_P\equiv \log\frac{M_{P}^2}{\bar{\mu}^2}$ for $P=0,K,\eta$, and $L_i^r$ are the NLO LECs of the $n_f=3$ chiral Lagrangian of~\cite{Gasser:1984gg}. The terms $\epsilon_2^M$ and $\epsilon_2^F$ are $\mathcal{O}(p^6)$ and depend on the LECs of the $n_f=3$ NNLO Lagrangian, $C_i^r$. Both $\epsilon_2^M$ and $\epsilon_2^F$ involve the calculation of a two-mass scale sunset integral at non-zero external momentum, and therefore do not admit a closed analytic expression (see~\cite{Amoros:1999dp}, where a two-integral representation is provided). However, by employing the analytic form of the two-mass scale sunset integral at vanishing external momentum~\cite{Davydychev:1992mt,Amoros:1999dp} and the recursion relations for sunset integrals~\cite{Tarasov:1997kx,Tarasov:1996br,Kaiser:2007kf}, $\epsilon_2^M$ and $\epsilon_2^F$ can be expanded in power series of $m$. Such an expansion has been calculated for the first time in~\cite{Kaiser:2007kf} for $\epsilon_2^M$ and~\cite{Ananthanarayan:2017yhz} for $\epsilon_2^F$ up to $\mathcal{O}(m^3)$ and $\mathcal{O}(m^2)$ respectively and is sufficient for the matching of the $n_f=2$ and $n_f=3$ LECs discussed in app.~\ref{app:capp}. Since the result of $\epsilon_2^M$ and $\epsilon_2^F$ turns out to be quite involved, we will avoid reporting here the explicit expressions, for which we refer to~\cite{Kaiser:2007kf,Ananthanarayan:2017yhz}.

Finally, the topological susceptibility at $\alpha_{\rm em}=0$ (first computed to $\mathcal{O}(p^4)$ in \cite{Mao:2009sy}) reads
\begin{align}\label{eq:topbare3f}
\chi_{\rm top} =   \frac{  m_um_d}{ m_u+m_d+\frac{m_um_d}{m_s}} B_0 F_0^2\left[1 + \epsilon_1^\chi+\epsilon_2^\chi\right] + \mathcal{O}(p^8) \, ,
\end{align}
where in the unbroken isospin limit $m_u=m_d\equiv m$:
\begin{align}
\epsilon_1^\chi =&\ \frac{B_0 m_s}{ F_0^2 (1+w/2)}\left\{\left[-\kappa\left(\frac{2 \lambda _{\eta }}{9}+ \lambda _K\right)+32 L_6^r\right]+\left[-\kappa\left(\frac{5  \lambda _{\eta }}{9}+3   \lambda _0+2  \lambda _K\right)+16 \left(5 L_6^r+9 L_7^r+3 L_8^r\right)\right]w\right.\nonumber\\ \,
&\left.\qquad\qquad\quad\ \ \, +\left[-\kappa\left(\frac{2 \lambda _{\eta }}{9}+ \lambda _K\right)+32 L_6^r\right]w^2\right\}\, , \\\label{eq:epsilonchi2}
\epsilon_2^\chi =&\ \frac{B_0^2 m_s^2}{ F_0^4(1+w/2)}\left[\epsilon_{2,C}^\chi+\epsilon_{2,\log\times\log}^\chi+\epsilon_{2,\log}^\chi+\epsilon_{2,\log\times L}^\chi+\epsilon_{2,L}^\chi+\epsilon_{2,L\times L}^\chi+\epsilon_{2,{\rm finite}}^\chi\right] \ .
\end{align}
The result of $\epsilon_2^\chi$ has been conveniently organized into different contributions: $\epsilon_{2,C}^\chi$ are terms containing the LECs $C_i^r$, $\epsilon_{2,\log}^\chi$ and $\epsilon_{2,\log\times\log}^\chi$ are terms respectively linear and quadratic in the chiral logs $\lambda_{0,K,\eta}$ (but without $L_i^r$), $\epsilon_{2,L}^\chi$ and $\epsilon_{2,L\times L}^\chi$ are terms linear and quadratic in $L_i^r$ (but independent of chiral logs), and $\epsilon_{2,\log\times L}^\chi$ contains products of chiral logs and LECs. Finally, $\epsilon_{2,{\rm finite}}^\chi$ is the remaining constant piece and is automatically scale independent. In particular:
{\small
\begin{align}
\epsilon_{2,C}^\chi =& \ 8\left\{ 8 \left[C_{20}^r+3 C_{21}^r\right]+4 \left[3 C_{19}^r+7 C_{20}^r+27 C_{21}^r+2 C_{31}^r+C_{94}^r+6 \left(C_{32}^r+C_{33}^r\right)\right]w+ \right. \\ \nonumber
&\ \left. 4 \left[6 C_{19}^r+C_{94}^r+4 \left(4 C_{20}^r+9 C_{21}^r+C_{31}^r+3 \left(C_{32}^r+C_{33}^r\right)\right)\right]w^2+ \left[8 C_{20}^r+48 C_{21}^r+C_{94}^r\right]w^3\right\}\, ,\\ \nonumber \\
\epsilon_{2,\log\times\log}^\chi = &\ \frac{\kappa^2}{2+w}\left\{\frac{4}{9}\left[ \lambda_\eta (\lambda_\eta-5 \lambda_K)\right]+\frac13\left[\frac{46 \lambda_\eta^2}{9}+\frac{47 \lambda_K^2}{3}-\frac{70 \lambda_\eta \lambda_K}{3}-4 \lambda_0^2+16 \lambda_\eta \lambda_0\right]w\right.\\\nonumber
&\left.+\left[10 \lambda_0 \lambda_{\eta}+\frac{5 \lambda_{\eta }^2}{3}-\frac{19\lambda _0^2}{3} -10 \lambda _{\eta} \lambda _K+18 \lambda _0 \lambda _K+\frac{14 \lambda _K^2}{9}\right]w^2\right.\\\nonumber
&\left.+\left[5 \lambda _0 \lambda _{\eta }+\frac{11 \lambda _{\eta }^2}{18}-\frac{71\lambda _0^2}{6}  -\frac{46 \lambda _{\eta } \lambda _K}{9}+12 \lambda _0 \lambda _K-\frac{7 \lambda _K^2}{9}\right]w^3+\frac13\left[\frac{2 \lambda _{\eta }^2}{9}+2 \lambda _0 \lambda _{\eta }-\frac{8 \lambda _{\eta } \lambda _K}{3}\right]w^4
\right\}\, ,\\ \nonumber\\
\epsilon_{2,\log}^\chi = &\ -\frac{\kappa^2}{3}\left\{\frac23\left[\frac{10 \lambda _{\eta }}{9}+ \lambda _K\right]+\frac23\left[\frac{29 \lambda _{\eta }}{3}+34 \lambda _K\right]w+2\left[\frac{19 \lambda _{\eta }}{9}+19 \lambda _0+\frac{35 \lambda _K}{3}\right]w^2\right.\\\nonumber
&\left.+2\left[\frac{8 \lambda _{\eta }}{27}-\lambda _0+\frac{2 \lambda _K}{3}\right]w^3
\right\}\, ,\\\nonumber\\
\epsilon_{2,\log\times L}^\chi = &\ \frac{8\kappa}{2+w}\left\{\left[\frac{16}{9} \left(3 L_4^r+L_5^r-6 L_6^r+3 L_7^r-L_8^r\right)\lambda_\eta+2 \left(8 L_4^r+3 L_5^r-16 L_6^r-6 L_8^r\right)\lambda_K\right]\right.\\\nonumber
&\left.+\bigg[\frac{8}{9} \left(18 L_4^r+7 L_5^r-36 L_6^r-27 L_7^r-23 L_8^r\right)\lambda_\eta+3 \left(20 L_4^r+7 L_5^r-40 L_6^r-24 L_7^r-22 L_8^r\right)\lambda_K\right.\\\nonumber
&\left.+24 \left(L_4^r-2 L_6^r\right)\lambda_0\bigg]w+\bigg[\frac{20}{3} \left(3 L_4^r+L_5^r-4 L_8^r-6 \left(L_6^r+L_7^r\right)\right)\lambda_\eta+\left(82 L_4^r+27 L_5^r-2 \left(82 L_6^r+72 L_7^r+51 L_8^r\right)\right)\lambda_K\right.\\\nonumber
&\left.+12 \left(7 L_4^r+3 L_5^r-2 \left(7 L_6^r+9 L_7^r+6 L_8^r\right)\right)\lambda_0\bigg]w^2+\bigg[\frac{2}{9} \left(48 L_4^r+13 L_5^r-96 L_6^r-60 L_7^r-46 L_8^r\right)\lambda_\eta\right.\\\nonumber
&+\left.\left(48 L_4^r+15 L_5^r-6 \left(16 L_6^r+12 L_7^r+9 L_8^r\right)\right)\lambda_K+6 \left(8 L_4^r+3 L_5^r-16 L_6^r-6 L_8^r\right)\lambda_0\bigg]w^3\right.\\\nonumber
&\left.+\left[\left(2L_4^r+\frac{4}{9} \left(L_5^r-9 L_6^r-2 L_8^r\right)\right)\lambda_\eta+\left(10 L_4^r+3 L_5^r-20 L_6^r-6 L_8^r\right)\lambda_K+6 \left(L_4^r-2 L_6^r\right)\lambda_0\right]w^4\right\}\, ,\\\nonumber
\\ 
\epsilon_{2, L}^\chi = &\ \frac{8\kappa}{3}\left\{\frac13\left[22 L_4^r+\frac{35 L_5^r}{3}-44 L_6^r-8 L_7^r-26 L_8^r\right]+\left[48 L_4^r+\frac{35 L_5^r}{3}-96 L_6^r-\frac{70 L_8^r}{3}\right]w\right.\\
&\left.+\left[74 L_4^r+29 L_5^r-148 L_6^r+8 L_7^r-\frac{166 L_8^r}{3}\right]w^2+\frac13\left[44 L_4^r+\frac{31 L_5^r}{3}-88 L_6^r-16 L_7^r-26 L_8^r\right]w^3\right\}\, ,\nonumber
\\ \nonumber \\
\epsilon_{2, L\times L}^\chi = &\ \frac{1024}{2+w}\left(3 L_7^r+L_8^r\right)^2\left\{-w+2w^2-w^3\right\}\, ,
\\\nonumber\\
\epsilon_{2, {\rm finite}}^\chi = & \ \kappa^2\left\{\left[-6  G\left(\frac{2 w}{w+1}\right)-\frac{2}{9}G\left(1\right)+\frac{2 }{3}G\left(\frac{2 + w}{3 w}\right)-\frac{4}{9}G\left(\frac{4}{3}\frac{1+w/2}{1+w}\right)+\frac{80}{9}\right]w\right.\\
&\left.\quad\,+\left[-3 G\left(\frac{2 w}{w+1}\right)-\frac{1}{9}G(1)-\frac{11}{3}G\left(\frac{2 + w}{3 w}\right)-\frac{5}{9}G\left(\frac{4}{3}\frac{1+w/2}{1+w}\right)+\frac{160}{9}\right]w^2\right\}\, .\nonumber
\end{align}}
In this last equation we defined
\begin{align}
G(x) \equiv \frac{1}{\sigma} \bigg[ 4 \text{Li}_2 \bigg( \frac{\sigma-1}{\sigma+1} \bigg) + \log^2 \bigg( \frac{1-\sigma}{1+\sigma} \bigg) + \frac{\pi^2}{3} \bigg] \, , \qquad \sigma = \sqrt{1-\frac{4}{x}}  \, ,
\end{align}
which arises in the evaluation of a two-mass scale sunset integral at vanishing external momentum~\cite{Ananthanarayan:2017yhz}. 

\section{Extraction of the NNLO LECs and input parameters}\label{app:capp}
As mentioned in sec.~\ref{sec:NNLOcorr}, the $\mathcal{O}(p^6)$ LECs $c_i^r$ of the $n_f=2$ chiral Lagrangian~\cite{Bijnens:1999sh,Bijnens:1999hw} are still poorly known from experimental data (see~\cite{Bijnens:2014lea} for a review). To estimate the value of the $c_i^r$'s appearing in eq.~(\ref{eq:delta2}), we combined the results from recent lattice QCD simulations~\cite{Boyle:2015exm} with the information from the matching of pion mass, decay constant and topological susceptibility for $n_f=2$ and $n_f=3$ chiral perturbation theory of app.~\ref{app:bareresults} and the estimates for the $n_f=3$ LECs $C_i^r$ provided in~\cite{Bijnens:2014lea}. 
\begin{itemize}
\item The SU(2) partially quenched simulation of ref.~\cite{Boyle:2015exm} provide fits of $8$ combinations of SU(2) LECs. In this analysis we consider the $450$ MeV cut-fit for such combinations, reported in the last column of tab.~6 of~\cite{Boyle:2015exm}. While this estimate is less conservative than the one extracted from the $370$~MeV cut-fit, the two are compatible for all reported combinations of LECs. In tab.~\ref{tab:clattice} we quote the results expressed in terms of $c_i^r$ using the relations between SU$(N)$ and SU($2$) LECs of~\cite{Bijnens:1999sh}.

\item Ref.~\cite{Bijnens:2014lea} provides estimates of $34$ combinations of $\mathcal{O}(p^6)$ three-flavor LECs, $C_i^r$. The information of $C_i^r$ can be translated into a value for three combinations of $c_i^r$ by equating in the large $m_s/m$ limit (and for $\alpha_{\rm em}=0$ and $m\equiv m_u=m_d$) the $n_f=2$ and $n_f=3$ formulas for the pion mass in eqs.~(\ref{eq:mpibare}) and (\ref{eq:mpibare3f}), the pion decay constant in eqs.~(\ref{eq:fpibare}) and (\ref{eq:fpibare3f}) and the topological susceptibility in eqs.~(\ref{eq:topbare}) and (\ref{eq:topbare3f}). Such a matching leads respectively to the three following relations:
\begin{table}[t]
\centering
\begin{tabular}{c|r}
\hline 
$ c_6^r-c_{17}^r $ & $ -5.33(77)\cdot 10^{-6}$\\ 
$ 2c_6^r-12c_{10}^r+18c_{11}^r-c_{18}^r $ & $ 14.5(3.9)\cdot 10^{-6}$  \\ 
$ c_7^r$ & $ -3.9(2.3)\cdot 10^{-6}$  \\ 
$ c_8^r$ & $ 0.0(1.8)\cdot 10^{-6}$  \\ 
$ 2c_7^r+4c_8^r$ & $ 6.2(3.2)\cdot 10^{-6}$  \\ 
$ c_9^r$ & $ -0.2(1.2)\cdot 10^{-6}$  \\ 
$ c_{10}^r$ & $ -1.0(1.1)\cdot 10^{-6}$  \\ 
$ 19c_{11}^r-12c_{10}^r$ & $ 10.1(3.1)\cdot 10^{-6}$  \\ \hline
\end{tabular}
\caption{\label{tab:clattice} \emph{Numerical values of the combinations of SU(2) LECs at the scale $\bar \mu=770$~MeV extracted from the $450$ MeV cut-fit of the partially quenched simulations in~\cite{Bijnens:2014lea} (tab.~6).}}
\end{table}

{\small\begin{align}\label{eq:CM}
&\quad2 c_6^r+c_7^r+2 c_8^r+c_9^r-3 c_{10}^r-2 \left(3 c_{11}^r+c_{17}^r+2 c_{18}^r\right)= \\\nonumber
&=\frac{\kappa f_\pi^2}{1152m_\pi^2}\frac{m}{m_s}+\left[2 C_{12}^r+4 C_{13}^r+C_{14}^r+2 C_{15}^r+2 C_{16}^r+C_{17}^r-3 C_{19}^r-2 \left(3 C_{20}^r+6 C_{21}^r+C_{31}^r+2 C_{32}^r\right)\right]\\\nonumber
&+	\frac{\kappa^2}{384}\left[\frac{1}{9}  \lambda _{\eta }^2+ \lambda _{\eta } \lambda _K+\frac{13}{6}  \lambda _K^2\right]+\frac{\kappa^2}{589824}\left[\frac{20975}{3}\lambda _{\eta }+44123 \lambda _K\right]+\kappa\bigg[\frac{1}{36} \left(4 L_1^r+L_2^r+L_3^r-10 L_4^r\right.\\\nonumber
&\left.-3 L_5^r+12 L_6^r+12 L_7^r+10 L_8^r\right)\lambda_\eta+\left(L_1^r+\frac{L_2^r}{4}+\frac{5 L_3^r}{16}-L_4^r+L_6^r+\frac{L_8^r}{2}-\frac{L_5^r}{4}\right)\lambda_K\bigg]+\\\nonumber
&+\frac{\kappa}{3}\left[5 L_1^r+\frac{5 L_2^r}{6}+\frac{205 L_3^r}{144}+6 L_6^r+\frac{11 L_7^r}{24}+\frac{73 L_8^r}{24}-\frac{17 L_4^r}{3}-\frac{13 L_5^r}{9}\right]\\\nonumber
&-4 \left[\left(2 L_4^r+L_5^r\right) \left(2 L_4^r+L_5^r-2 \left(2 L_6^r+L_8^r\right)\right)\right]+\frac{\kappa^2}{589824}\left[\frac{1373}{4} G\left(\frac{4}{3}\right)+\frac{219836}{3}\right] \, ,\\\nonumber\\\label{eq:CF}
&\quad c_7^r+2 c_8^r+c_9^r=\\\nonumber
&=-\frac{\kappa f_\pi^2}{64m_\pi^2}\frac{m}{m_s}+\left[C_{14}^r+2 C_{15}^r+2 C_{16}^r+C_{17}^r\right]-\frac{\kappa^2}{384}\left[\frac{7}{3} \lambda _{\eta } \lambda _K+ \lambda _K^2\right]+\frac{\kappa^2}{589824}\left[44549  \lambda _K-10245  \lambda _{\eta }\right]\\\nonumber
&+\kappa\left[\frac{1}{36}\left(4 L_1^r+L_2^r+L_3^r-2 L_4^r-L_5^r\right)\lambda_\eta+\left(L_1^r+\frac{L_2^r}{4}+\frac{5 L_3^r}{16}+\frac{L_4^r}{4}-L_6^r-\frac{L_8^r}{4}\right)\lambda_K\right]+\\\nonumber
&+\frac{\kappa}{3}\left[5 L_1^r+\frac{5 L_2^r}{6}+\frac{205 L_3^r}{144}+\frac{5 L_4^r}{4}+\frac{23 L_5^r}{48}-6 L_6^r-\frac{15 L_8^r}{8}\right]-\left[2 L_4^r+L_5^r\right]^2+\frac{\kappa^2}{49152}\left[\frac{5825}{16} G\left(\frac{4}{3}\right)+5225\right]\, ,\\ \nonumber\\\label{eq:CA}
&\quad 3 c_{10}^r+6 c_{11}^r+2 c_{17}^r+4 c_{18}^r+4 c_{19}^r=\\\nonumber
&=\frac{f_\pi^2}{2m_\pi^2}\frac{m}{m_s}\left[\frac{\kappa}{32}\left(\frac{\lambda _{\eta }}{3}+\frac{\lambda _K}{2}-\frac{23}{18}\right)-\left(\frac{3L_{6}^r}{2} +9 L_{7}^r+3 L_{8}^r\right)\right]+\bigg[\frac{3 C_{19}^r}{2}+\frac{21 C_{20}^r}{4}+\frac{27 C_{21}^r}{4}+C_{31}^r\\\nonumber
&+3 C_{32}^r+3 C_{33}^r\bigg]-\frac{\kappa^2}{27}\left[\lambda _{\eta }^2+\frac{3\lambda _K}{256}  \left(26 \lambda _{\eta }+113 \lambda _K\right)\right]-\frac{\kappa^2}{768}\left[\frac{1346 \lambda _{\eta }}{27}+95 \lambda _K\right]+\frac{\kappa}{4}\left[\left(\frac{4 L_4^r}{3}-3 L_7^r-L_8^r\right.\right.\\\nonumber
&\left.\left.-\frac{10 L_6^r}{3}\right)\lambda_\eta+\frac{3}{8} \left(12 L_4^r+3 L_5^r-24 L_6^r-24 L_7^r-14 L_8^r\right)\lambda_K\right]+\frac{\kappa}{24}\left[\frac{317 L_4^r}{6}+\frac{217 L_5^r}{12}-127 L_7^r-\frac{157 L_8^r}{2}\right.\\\nonumber
&\left.-\frac{317 L_6^r}{3}\right]+4 \left[-3 \left(L_6^r+6 L_7^r+2 L_8^r\right) L_4^r+6 \left(L_6^{r}\right)^2+7 \left(3 L_7^r+L_8^r\right){}^2+12 L_6^r \left(3 L_7^r+L_8^r\right)\right]+\frac{\kappa ^2}{576}\bigg[G(1)\\\nonumber
&+\frac{1}{2} G\left(\frac{4}{3}\right)-7 \pi ^2-\frac{2275}{9}\bigg] \, .
\end{align}}

The notation in eqs.~(\ref{eq:CM}-\ref{eq:CA}) is as in app.~\ref{subsec:3flavbare} (except that
 $m/m_\pi^2$, $f_\pi$ and $M^2_{K,\eta}$ in $\lambda_{K,\eta}$ are computed at $m_u=m_d=m=0$),
 and the contributions on the r.h.s. have been ordered as in eq.~(\ref{eq:epsilonchi2}). The numerical value of the NLO couplings $L_i^r$ is reported in tab.~\ref{tab:LCs}: in particular, $L_{4}^r,L_{5}^r$ and $L_{6}^r$ are taken from lattice QCD studies~\cite{Aoki:2016frl}, while the others from ref.~\cite{Bijnens:2014lea}. Moreover, to all $L_i^r$ a 30\% intrinsic uncertainty from higher order 3-flavor corrections has been added (this is not present for 2-flavor where higher order corrections are much smaller). The value of the NNLO couplings $C_i^r$ appearing in the r.h.s. of eqs.~(\ref{eq:CM}-\ref{eq:CA}) taken from tab.~4 of ref.~\cite{Bijnens:2014lea} is also reported in tab.~\ref{tab:LCs}. 
Since ref.~\cite{Bijnens:2014lea} did not provide uncertainties for the $C_i^r$ coefficients we assume that they reproduce at least the right orders of magnitude and conservatively assign to them a $100\%$ uncertainty.

Eqs.~(\ref{eq:CM}-\ref{eq:CA}) then lead to:
\begin{align}\nonumber
2 c_6^r+c_7^r+2 c_8^r+c_9^r-3 c_{10}^r-2 \left(3 c_{11}^r+c_{17}^r+2 c_{18}^r\right)=& -3.5(22.0)\cdot 10^{-6}\, ,\label{eq:Cnum}\\
 c_7^r+2 c_8^r+c_9^r=& \ 4.7(9.2)\cdot 10^{-6} \, ,\\\nonumber
 3 c_{10}^r+6 c_{11}^r+2 c_{17}^r+4 c_{18}^r+4 c_{19}^r=&\ 0.3(25.5)\cdot 10^{-6}\, .
\end{align}
\end{itemize}

\begin{table}[t]
\centering
\begin{tabular}{cccccccc|c}
\hline
$ L_1^r$ & $ L_2^r$ & $  L_3^r$ & $  L_4^r$ & $ L_5^r$ & $  L_6^r$ & $  L_7^r$ & $  L_8^r$ & \\ \hline
$ 0.5(2)$ & $ 0.8(3)$ & $ -3.1(1.0)$ & $ 0.09(34)$ & $ 1.19(40)$ & $ 0.16(20)$ & $ -0.34(11)$ & $ 0.55(18)$ & $\times 10^{-3}$\\ \hline
\end{tabular}\\ \vspace{4mm}
\centering
\begin{tabular}{cccccccccccc|c}
\hline 
$ C_{12}^r$ & $  C_{13}^r$   & $  C_{14}^r$ & $  C_{15}^r$ & $  C_{16}^r$ & $  C_{17}^r$  &
$  C_{19}^r$& $  C_{20}^r$& $  C_{21}^r$& $  C_{31}^r$& $  C_{32}^r$& $  C_{33}^r$\\ \hline
$ -2.8$ & $ 1.5$ & $ -1.0$ & $ -3.0$ & $ 3.2$ & $ -1.0$  & 
$ -4.0$& $1.0$& $-0.48$& $ 2.0$& $1.7$& $ 0.82$ & $\times 10^{-6}$\\ \hline
\end{tabular}
\caption{\label{tab:LCs} \emph{Numerical value of the NLO and NNLO couplings $L_i^r$ and $C_i^r$ at the scale $\bar \mu=770$~MeV. We associated $100\%$ uncertainty to the $C_i^r$.}}
\end{table}
The final value of the $9$ couplings $c_i^r$ in tab.~\ref{tab:cs} has been extracted by combining the lattice results in tab.~\ref{tab:clattice} with the 2-3 flavor matching result in eq.~(\ref{eq:Cnum}) through a $\chi^2$ fit, whose quality ($\chi^2\sim3$) turns out to be good. In principle, an estimate of $c_6^r$ could be directly extracted from the pion scalar form factor, as in ref.~\cite{Bijnens:2014lea} where $c_6^r\approx-1.9\times10^{-5}$. However, since there is still a factor $\sim3$ uncertainty on how to theoretically model this last quantity~\cite{Donoghue:1990xh,Bijnens:1998fm,Caprini:2018jjm}, we chose not to use this estimate of $c_6^r$ in our numerical analysis. In any case, the NNLO corrections to $\chi_{\rm top}$ in eq.~(\ref{eq:delta2}) taking into account also $c_6^r=-1.9(1.9)\times10^{-5}$ result in $\delta_2=-0.006(3)$, still compatible with eq.~(\ref{eq:delta2num}), but with an overall lower quality fit of the $c_i^r$.

Finally, for convenience in tab.~\ref{tab:numval} we summarize the values of the parameters used in this work, which should be considered together with the LECs in tabs.~\ref{tab:ks}, \ref{tab:cbarKXs}, \ref{tab:cs} and \ref{tab:LCs}. When uncertainties are not quoted it means that their effect was negligible and they have not been used.

\begin{table}[t]
\centering
\begin{tabular}{c c c| c c c }
\hline  
$z$ & 0.472(11) & eq. (\ref{eq:znow})                & $w^{-1}$ & 27 & \cite{Aoki:2016frl} \\ \hline
$f_\pi$ & 92.3 & \cite{Tanabashi:2018oca} & $\bar{\ell}_3$ & 2.81(49) &\cite{Boyle:2015exm}  \\ \hline 
$m_{\pi^0}$ & 134.98 & \cite{Tanabashi:2018oca} & $\bar{\ell}_4$ & 4.02(25) &\cite{Boyle:2015exm}  \\ \hline
$m_{\pi^+}$ & 139.57 & \cite{Tanabashi:2018oca} & $h_1^r-h_3^r-\ell_4^r$ & -0.0049(12) & eq. (\ref{eq:htl7}) \\ \hline
$m_K$ & $495$ & \cite{Tanabashi:2018oca} & $\ell_7$ & 0.0065(38) &\cite{Boyle:2015exm} \\ \hline
$m_\rho$ & 775 & \cite{Tanabashi:2018oca} &$\alpha_{\rm em}^{-1}$ & $137$ & \cite{Tanabashi:2018oca} \\ \hline
$m_\mu$ & 105.658 &\cite{Tanabashi:2018oca}  & $\Gamma_{\pi^{+}\to\mu\nu(\gamma)}$ & $2.5281 \cdot 10^{-14}$ & \cite{Tanabashi:2018oca}\\ \hline
$G_F$ & $1.16638\cdot 10^{-11}$ & \cite{Tanabashi:2018oca} & $V_{ud}$ & $0.9742$ 
&\cite{Tanabashi:2018oca} 
\\ \hline
\end{tabular}
\caption{\label{tab:numval} \emph{Numerical input values used in the computations. 
Dimensionful quantities are given in MeV.}}
\end{table}



\begin{thebibliography}{99}


\bibitem{Berkowitz:2015aua}
  E.~Berkowitz, M.~I.~Buchoff and E.~Rinaldi,
  ``Lattice QCD input for axion cosmology,''
  Phys.\ Rev.\ D {\bf 92} (2015) no.3,  034507
  doi:10.1103/PhysRevD.92.034507
  [arXiv:1505.07455 [hep-ph]].
  
\bibitem{Borsanyi:2015cka}
  S.~Borsanyi {\it et al.},
  ``Axion cosmology, lattice QCD and the dilute instanton gas,''
  Phys.\ Lett.\ B {\bf 752} (2016) 175
  doi:10.1016/j.physletb.2015.11.020
  [arXiv:1508.06917 [hep-lat]].
  
\bibitem{Petreczky:2016vrs}
  P.~Petreczky, H.~P.~Schadler and S.~Sharma,
  ``The topological susceptibility in finite temperature QCD and axion cosmology,''
  Phys.\ Lett.\ B {\bf 762} (2016) 498
  doi:10.1016/j.physletb.2016.09.063
  [arXiv:1606.03145 [hep-lat]].


\bibitem{Borsanyi:2016ksw}
  S.~Borsanyi {\it et al.},
  ``Calculation of the axion mass based on high-temperature lattice quantum chromodynamics,''
  Nature {\bf 539} (2016) no.7627,  69
  doi:10.1038/nature20115
  [arXiv:1606.07494 [hep-lat]].

\bibitem{Burger:2018fvb}
  F.~Burger, E.~M.~Ilgenfritz, M.~P.~Lombardo and A.~Trunin,
  ``Chiral observables and topology in hot QCD with two families of quarks,''
  Phys.\ Rev.\ D {\bf 98} (2018) no.9,  094501
  doi:10.1103/PhysRevD.98.094501
  [arXiv:1805.06001 [hep-lat]].

\bibitem{Bonati:2018blm}
  C.~Bonati, M.~D'Elia, G.~Martinelli, F.~Negro, F.~Sanfilippo and A.~Todaro,
  ``Topology in full QCD at high temperature: a multicanonical approach,''
  arXiv:1807.07954 [hep-lat].

  
\bibitem{Peccei:1977hh}
  R.~D.~Peccei and H.~R.~Quinn,
  ``CP Conservation in the Presence of Instantons,''
  Phys.\ Rev.\ Lett.\  {\bf 38} (1977) 1440.
  
  
\bibitem{Weinberg:1977ma}
  S.~Weinberg,
  ``A New Light Boson?,''
  Phys.\ Rev.\ Lett.\  {\bf 40} (1978) 223.

\bibitem{Wilczek:1977pj}
  F.~Wilczek,
  ``Problem of Strong p and t Invariance in the Presence of Instantons,''
  Phys.\ Rev.\ Lett.\  {\bf 40} (1978) 279.

\bibitem{Kim:1979if}
  J.~E.~Kim,
  ``Weak Interaction Singlet and Strong CP Invariance,''
  Phys.\ Rev.\ Lett.\  {\bf 43} (1979) 103.
  
\bibitem{Shifman:1979if}
  M.~A.~Shifman, A.~I.~Vainshtein and V.~I.~Zakharov,
  ``Can Confinement Ensure Natural CP Invariance of Strong Interactions?,''
  Nucl.\ Phys.\ B {\bf 166} (1980) 493.

\bibitem{Zhitnitsky:1980tq}
  A.~R.~Zhitnitsky,
  ``On Possible Suppression of the Axion Hadron Interactions. (In Russian),''
  Sov.\ J.\ Nucl.\ Phys.\  {\bf 31} (1980) 260
   [Yad.\ Fiz.\  {\bf 31} (1980) 497].
    
\bibitem{Dine:1981rt}
  M.~Dine, W.~Fischler and M.~Srednicki,
  ``A Simple Solution to the Strong CP Problem with a Harmless Axion,''
  Phys.\ Lett.\ B {\bf 104} (1981) 199.

\bibitem{Tanabashi:2018oca}
  M.~Tanabashi {\it et al.} [Particle Data Group],
  ``Review of Particle Physics,''
  Phys.\ Rev.\ D {\bf 98} (2018) no.3,  030001.
  doi:10.1103/PhysRevD.98.030001



\bibitem{Preskill:1982cy}
  J.~Preskill, M.~B.~Wise and F.~Wilczek,
  ``Cosmology of the Invisible Axion,''
  Phys.\ Lett.\ B {\bf 120} (1983) 127.

\bibitem{Abbott:1982af}
  L.~F.~Abbott and P.~Sikivie,
  ``A Cosmological Bound on the Invisible Axion,''
  Phys.\ Lett.\ B {\bf 120} (1983) 133.

\bibitem{Dine:1982ah}
  M.~Dine and W.~Fischler,
  ``The Not So Harmless Axion,''
  Phys.\ Lett.\ B {\bf 120} (1983) 137.
    

\bibitem{Graham:2015ouw}
  P.~W.~Graham, I.~G.~Irastorza, S.~K.~Lamoreaux, A.~Lindner and K.~A.~van Bibber,
  ``Experimental Searches for the Axion and Axion-Like Particles,''
  Ann.\ Rev.\ Nucl.\ Part.\ Sci.\  {\bf 65} (2015) 485
  doi:10.1146/annurev-nucl-102014-022120
  [arXiv:1602.00039 [hep-ex]].

\bibitem{Irastorza:2018dyq}
  I.~G.~Irastorza and J.~Redondo,
  ``New experimental approaches in the search for axion-like particles,''
  arXiv:1801.08127 [hep-ph].



\bibitem{diCortona:2015ldu}
  G.~Grilli di Cortona, E.~Hardy, J.~Pardo Vega and G.~Villadoro,
  ``The QCD axion, precisely,''
  JHEP {\bf 1601} (2016) 034
  doi:10.1007/JHEP01(2016)034
  [arXiv:1511.02867 [hep-ph]].


\bibitem{Gasser:1983yg}
  J.~Gasser and H.~Leutwyler,
  ``Chiral Perturbation Theory to One Loop,''
  Annals Phys.\  {\bf 158} (1984) 142.
  doi:10.1016/0003-4916(84)90242-2

\bibitem{Bonati:2015vqz}
  C.~Bonati, M.~D'Elia, M.~Mariti, G.~Martinelli, M.~Mesiti, F.~Negro, F.~Sanfilippo and G.~Villadoro,
  ``Axion phenomenology and $\theta$-dependence from $N_f = 2+1$ lattice QCD,''
  JHEP {\bf 1603} (2016) 155
  doi:10.1007/JHEP03(2016)155
  [arXiv:1512.06746 [hep-lat]].


\bibitem{Fodor:2016bgu}
  Z.~Fodor {\it et al.},
  ``Up and down quark masses and corrections to Dashen's theorem from lattice QCD and quenched QED,''
  Phys.\ Rev.\ Lett.\  {\bf 117} (2016) no.8,  082001
  doi:10.1103/PhysRevLett.117.082001
  [arXiv:1604.07112 [hep-lat]].

\bibitem{Giusti:2017dmp}
  D.~Giusti, V.~Lubicz, C.~Tarantino, G.~Martinelli, S.~Sanfilippo, S.~Simula and N.~Tantalo,
  ``Leading isospin-breaking corrections to pion, kaon and charmed-meson masses with Twisted-Mass fermions,''
  Phys.\ Rev.\ D {\bf 95} (2017) no.11,  114504
  doi:10.1103/PhysRevD.95.114504
  [arXiv:1704.06561 [hep-lat]].

\bibitem{Basak:2018yzz}
  S.~Basak {\it et al.} [MILC Collaboration],
  ``Lattice computation of the electromagnetic contributions to kaon and pion masses,''
  arXiv:1807.05556 [hep-lat].

\bibitem{Carrasco:2014cwa}
  N.~Carrasco {\it et al.} [European Twisted Mass Collaboration],
  ``Up, down, strange and charm quark masses with N$_f$ = 2+1+1 twisted mass lattice QCD,''
  Nucl.\ Phys.\ B {\bf 887} (2014) 19
  doi:10.1016/j.nuclphysb.2014.07.025
  [arXiv:1403.4504 [hep-lat]].

\bibitem{Aoki:2016frl}
  S.~Aoki {\it et al.},
  ``Review of lattice results concerning low-energy particle physics,''
  Eur.\ Phys.\ J.\ C {\bf 77} (2017) no.2,  112
  doi:10.1140/epjc/s10052-016-4509-7
  [arXiv:1607.00299 [hep-lat]].

\bibitem{Boyle:2015exm}
  P.~A.~Boyle {\it et al.},
  ``Low energy constants of SU(2) partially quenched chiral perturbation theory from N$_f$=2+1 domain wall QCD,''
  Phys.\ Rev.\ D {\bf 93} (2016) no.5,  054502
  doi:10.1103/PhysRevD.93.054502
  [arXiv:1511.01950 [hep-lat]].


  



\bibitem{Gasser:1982ap}
  J.~Gasser and H.~Leutwyler,
  ``Quark Masses,''
  Phys.\ Rept.\  {\bf 87} (1982) 77.
  doi:10.1016/0370-1573(82)90035-7

\bibitem{Knecht:1997jw}
  M.~Knecht and R.~Urech,
  ``Virtual photons in low-energy pi pi scattering,''
  Nucl.\ Phys.\ B {\bf 519} (1998) 329
  doi:10.1016/S0550-3213(98)00044-3
  [hep-ph/9709348].

\bibitem{Haefeli:2007ey}
  C.~Haefeli, M.~A.~Ivanov and M.~Schmid,
  ``Electromagnetic low-energy constants in chiPT,''
  Eur.\ Phys.\ J.\ C {\bf 53} (2008) 549
  doi:10.1140/epjc/s10052-007-0493-2
  [arXiv:0710.5432 [hep-ph]].

\bibitem{Ananthanarayan:2004qk}
  B.~Ananthanarayan and B.~Moussallam,
  ``Four-point correlator constraints on electromagnetic chiral parameters and resonance effective Lagrangians,''
  JHEP {\bf 0406} (2004) 047
  doi:10.1088/1126-6708/2004/06/047
  [hep-ph/0405206].

\bibitem{Bijnens:1996kk}
  J.~Bijnens and J.~Prades,
  ``Electromagnetic corrections for pions and kaons: Masses and polarizabilities,''
  Nucl.\ Phys.\ B {\bf 490} (1997) 239
  doi:10.1016/S0550-3213(97)00107-7
  [hep-ph/9610360].

\bibitem{Cirigliano:2007ga}
  V.~Cirigliano and I.~Rosell,
  ``pi/K $\to$ e anti-nu(e) branching ratios to O($e^2 p^4$) in Chiral Perturbation Theory,''
  JHEP {\bf 0710} (2007) 005
  doi:10.1088/1126-6708/2007/10/005
  [arXiv:0707.4464 [hep-ph]].

\bibitem{DescotesGenon:2005pw}
  S.~Descotes-Genon and B.~Moussallam,
  ``Radiative corrections in weak semi-leptonic processes at low energy: A Two-step matching determination,''
  Eur.\ Phys.\ J.\ C {\bf 42} (2005) 403
  doi:10.1140/epjc/s2005-02316-8
  [hep-ph/0505077].

\bibitem{Urech:1994hd}
  R.~Urech,
  ``Virtual photons in chiral perturbation theory,''
  Nucl.\ Phys.\ B {\bf 433} (1995) 234
  doi:10.1016/0550-3213(95)90707-N
  [hep-ph/9405341].

\bibitem{Knecht:1999ag}
  M.~Knecht, H.~Neufeld, H.~Rupertsberger and P.~Talavera,
  ``Chiral perturbation theory with virtual photons and leptons,''
  Eur.\ Phys.\ J.\ C {\bf 12} (2000) 469
  doi:10.1007/s100529900265
  [hep-ph/9909284].

\bibitem{Lubicz:2016mpj}
  V.~Lubicz, G.~Martinelli, C.~T.~Sachrajda, F.~Sanfilippo, S.~Simula, N.~Tantalo and C.~Tarantino,
  ``Electromagnetic corrections to the leptonic decay rates of charged pseudoscalar mesons: lattice results,''
  PoS LATTICE {\bf 2016} (2016) 290
  doi:10.22323/1.256.0290
  [arXiv:1610.09668 [hep-lat]].


\bibitem{Giusti:2017dwk}
  D.~Giusti, V.~Lubicz, G.~Martinelli, C.~T.~Sachrajda, F.~Sanfilippo, S.~Simula, N.~Tantalo and C.~Tarantino,
  ``First lattice calculation of the QED corrections to leptonic decay rates,''
  Phys.\ Rev.\ Lett.\  {\bf 120} (2018) no.7,  072001
  doi:10.1103/PhysRevLett.120.072001
  [arXiv:1711.06537 [hep-lat]].

\bibitem{Bijnens:1999sh}
  J.~Bijnens, G.~Colangelo and G.~Ecker,
  ``The Mesonic chiral Lagrangian of order $p^6$,''
  JHEP {\bf 9902} (1999) 020
  doi:10.1088/1126-6708/1999/02/020
  [hep-ph/9902437].



\bibitem{Bijnens:1999hw}
  J.~Bijnens, G.~Colangelo and G.~Ecker,
  ``Renormalization of chiral perturbation theory to order $p^6$,''
  Annals Phys.\  {\bf 280} (2000) 100
  doi:10.1006/aphy.1999.5982
  [hep-ph/9907333].

\bibitem{Bijnens:1998fm}
  J.~Bijnens, G.~Colangelo and P.~Talavera,
  ``The Vector and scalar form-factors of the pion to two loops,''
  JHEP {\bf 9805} (1998) 014
  doi:10.1088/1126-6708/1998/05/014
  [hep-ph/9805389].

\bibitem{Donoghue:1990xh}
  J.~F.~Donoghue, J.~Gasser and H.~Leutwyler,
  ``The Decay of a Light Higgs Boson,''
  Nucl.\ Phys.\ B {\bf 343} (1990) 341.
  doi:10.1016/0550-3213(90)90474-R

\bibitem{Caprini:2018jjm}
  I.~Caprini,
  ``Model-independent constraint on the pion scalar form factor and light quark masses,''
  Phys.\ Rev.\ D {\bf 98} (2018) no.5,  056008
  doi:10.1103/PhysRevD.98.056008
  [arXiv:1803.04150 [hep-ph]].

\bibitem{Bijnens:2014lea}
  J.~Bijnens and G.~Ecker,
  ``Mesonic low-energy constants,''
  Ann.\ Rev.\ Nucl.\ Part.\ Sci.\  {\bf 64} (2014) 149
  doi:10.1146/annurev-nucl-102313-025528
  [arXiv:1405.6488 [hep-ph]].

\bibitem{Ananthanarayan:2017yhz}
  B.~Ananthanarayan, J.~Bijnens and S.~Ghosh,
  ``An Analytic Analysis of the Pion Decay Constant in Three-Flavoured Chiral Perturbation Theory,''
  Eur.\ Phys.\ J.\ C {\bf 77} (2017) no.7,  497
  doi:10.1140/epjc/s10052-017-5019-y
  [arXiv:1703.00141 [hep-ph]].




\bibitem{Amoros:1999dp}
  G.~Amoros, J.~Bijnens and P.~Talavera,
  ``Two point functions at two loops in three flavor chiral perturbation theory,''
  Nucl.\ Phys.\ B {\bf 568} (2000) 319
  doi:10.1016/S0550-3213(99)00674-4
  [hep-ph/9907264].


\bibitem{Burgi:1996qi}
  U.~Burgi,
  ``Charged pion pair production and pion polarizabilities to two loops,''
  Nucl.\ Phys.\ B {\bf 479} (1996) 392
  doi:10.1016/0550-3213(96)00454-3
  [hep-ph/9602429].

\bibitem{Bijnens:1997vq}
  J.~Bijnens, G.~Colangelo, G.~Ecker, J.~Gasser and M.~E.~Sainio,
  ``Pion-pion scattering at low energy,''
  Nucl.\ Phys.\ B {\bf 508} (1997) 263
   Erratum: [Nucl.\ Phys.\ B {\bf 517} (1998) 639]
  doi:10.1016/S0550-3213(97)80013-2, 10.1016/S0550-3213(97)00621-4, 10.1016/S0550-3213(98)00127-8
  [hep-ph/9707291].


\bibitem{Colangelo:2001df}
  G.~Colangelo, J.~Gasser and H.~Leutwyler,
  ``$\pi \pi$ scattering,''
  Nucl.\ Phys.\ B {\bf 603} (2001) 125
  doi:10.1016/S0550-3213(01)00147-X
  [hep-ph/0103088].
  
\bibitem{Bijnens:2017wba}
  J.~Bijnens and N.~Hermansson Truedsson,
  ``The Pion Mass and Decay Constant at Three Loops in Two-Flavour Chiral Perturbation Theory,''
  JHEP {\bf 1711} (2017) 181
  doi:10.1007/JHEP11(2017)181
  [arXiv:1710.01901 [hep-ph]].
 
\bibitem{Mao:2009sy}
  Y.~Y.~Mao {\it et al.} [TWQCD Collaboration],
  ``Topological Susceptibility to the One-Loop Order in Chiral Perturbation Theory,''
  Phys.\ Rev.\ D {\bf 80} (2009) 034502
  doi:10.1103/PhysRevD.80.034502
  [arXiv:0903.2146 [hep-lat]].
 
  
  
\bibitem{Gasser:1984gg}
  J.~Gasser and H.~Leutwyler,
  ``Chiral Perturbation Theory: Expansions in the Mass of the Strange Quark,''
  Nucl.\ Phys.\ B {\bf 250} (1985) 465.
  doi:10.1016/0550-3213(85)90492-4
  

\bibitem{Davydychev:1992mt}
  A.~I.~Davydychev and J.~B.~Tausk,
  ``Two loop selfenergy diagrams with different masses and the momentum expansion,''
  Nucl.\ Phys.\ B {\bf 397} (1993) 123.
  doi:10.1016/0550-3213(93)90338-P
  
 
\bibitem{Kaiser:2007kf}
  R.~Kaiser,
  ``On the two-loop contributions to the pion mass,''
  JHEP {\bf 0709} (2007) 065
  doi:10.1088/1126-6708/2007/09/065
  [arXiv:0707.2277 [hep-ph]].

\bibitem{Tarasov:1996br}
  O.~V.~Tarasov,
  ``Connection between Feynman integrals having different values of the space-time dimension,''
  Phys.\ Rev.\ D {\bf 54} (1996) 6479
  doi:10.1103/PhysRevD.54.6479
  [hep-th/9606018].


\bibitem{Tarasov:1997kx}
  O.~V.~Tarasov,
  ``Generalized recurrence relations for two loop propagator integrals with arbitrary masses,''
  Nucl.\ Phys.\ B {\bf 502} (1997) 455
  doi:10.1016/S0550-3213(97)00376-3
  [hep-ph/9703319].




\end{thebibliography}
\end{document}